\documentclass[11pt]{article}

\usepackage{amsmath}
\usepackage{amssymb}
\usepackage{bbm}
\usepackage{cancel}
\usepackage{mathtools}
\usepackage{slashed}
\usepackage[numbers,sort&compress]{natbib}

\usepackage[letterpaper,margin=1in,bottom=1in]{geometry}
\usepackage{float} 
\usepackage{parskip} 
\usepackage{tabulary} 
\usepackage{color} 
\usepackage{soul} 
\usepackage{subfigure}
\usepackage{graphicx}
\usepackage[section]{placeins} 

\usepackage{cleveref}

\newcommand{\code}[1]{\texttt{#1}}
\newcommand{\GeV}{~\textrm{GeV}}

\newcommand{\gappeq}{\mathrel{\rlap {\raise.5ex\hbox{$>$}}
{\lower.5ex\hbox{$\sim$}}}}
\newcommand{\lappeq}{\mathrel{\rlap{\raise.5ex\hbox{$<$}}
{\lower.5ex\hbox{$\sim$}}}}

\author{
Bryan Kaufman\footnote{Email:b.kaufman@neu.edu},
Pran Nath\footnote{Email:p.nath@neu.edu},
Brent D Nelson\footnote{Email:b.nelson@neu.edu},~and
Andrew B. Spisak\footnote{Email:a.spisak@neu.edu}
\\~\\
Department of Physics, Northeastern University,
Boston, MA 02115-5000, USA
}

\title{
Light Stops and Observation of Supersymmetry at LHC RUN-II}

\begin{document}
\maketitle
\date

\textbf{Abstract: }
Light stops consistent with the Higgs boson mass of $\sim126\,{\rm GeV}$ are investigated within the framework of minimal supergravity. It is shown that models with light stops which are also consistent with the thermal relic density constraints require stop coannihilation with the neutralino LSP. The analysis shows that the residual set of parameter points with light stops satisfying both the Higgs mass and the relic density constraints lie within a series of thin strips in the $m_0-m_{1/2}$ plane for different values of $A_0/m_0$. Consequently, this region of minimal supergravity parameter space makes a number of very precise predictions. 
It is found that light stops of mass down to 400~GeV or lower can exist consistent with all constraints. A signal analysis for  this class  of models at  LHC RUN-II is carried out and the dominant signals for their detection identified. Also computed is the minimum integrated luminosity for $5\sigma$ discovery of the models analyzed. If supersymmetry is realized in this manner, the stop masses can be as low as 400~GeV or lower, and the mass gap between the lightest neutralino and lightest stop will be approximately 30-40~GeV. We have optimized the ATLAS signal regions specifically for stop searches in the parameter space and find that a stop with mass $\sim 375\,{\rm GeV}$ can be discovered with as little as $\sim$ 60~fb$^{-1}$ of integrated luminosity at RUN-II of the LHC; the integrated luminosity needed for discovery could be further reduced with more efficient signature analyses. The direct detection of dark matter in this class of models is also discussed. It is found that dark matter cross sections lie close to, but above, coherent neutrino scattering and would require multi-ton detectors such as LZ to see a signal of dark matter for this class of models.

\section{Introduction}\label{sec:intro}
The discovery of the Higgs boson in 2012~\cite{Chatrchyan:2012ufa, Aad:2012tfa} was a remarkable success of the Standard Model and established that the Higgs boson is indeed responsible for 
the breaking of the electroweak symmetry~\cite{Englert:1964et, Higgs:1964pj, Guralnik:1964eu}. In the context of supersymmetry (for a review see~\cite{Nath:2010zj}), the observed Higgs boson can be identified as the light $CP$-even state $h^0$ (see~\cite{Akula:2011aa, Baer:2011ab, Arbey:2011ab, Draper:2011aa, Carena:2011aa, Akula:2012kk, Arbey:2012dq, Strege:2012bt, Buchmueller:2011ab}). Further, the fact that the observed Higgs boson mass is $\sim126\GeV$ implies the need for a large loop correction to its mass to raise its tree value (which lies below $M_Z$) to the desired experimentally observed value. A large loop correction in turn implies a large supersymmetry-breaking weak scale on the order of a few TeV. However, a high SUSY scale does not preclude  some of the sparticle masses lying much lower in a widely split sparticle mass spectrum. In this work we focus on the possibility that, in spite of the overall SUSY scale being high, there might exist light squarks, specifically light stops, which may be discoverable at LHC RUN-II.

In models with large trilinear couplings, a large split between the stop masses is automatic, and for the case when the mass gap between the light stop and the LSP is small, i.e.,
\begin{align}
    (m_{\tilde{t}_1}-m_{\tilde{\chi}_1^0}) << m_{\tilde{\chi}_1^0}
    \label{eq:1}
\end{align}
the decay of the stop is likely to produce soft jets, making detection more difficult. This possibility becomes a requirement if the model is constrained to satisfy WMAP~\cite{Larson:2010gs} and Planck~\cite{Ade:2015xua} relic density constraints. Thus a satisfaction of the WMAP and Planck relic density constraints in this case requires that the lightest stop mass be within 20\% or less of the LSP mass (for some recent works on light stops see {~\cite{Papucci:2011wy, Bi:2011ha, Desai:2011th, Dutta:2013gga, Barnard:2014joa, Belanger:2015vwa, Hikasa:2015lma, Crivellin:2015bva, Baer:2015fsa,Chakraborty:2015wga,Demir:2014jqa}).
In the analysis of the light stops one must take account of the constraint that there be no instability arising from color and charge breaking 
minima. For most recent works on the necessary constraints for stability see ~\cite{Camargo-Molina:2013sta,Camargo-Molina:2014pwa}.}\\

We will work within the framework of supergravity grand unification~\cite{Chamseddine:1982jx, Nath:1983aw, Hall:1983iz, Arnowitt:1992aq} with radiative breaking of the electroweak symmetry (for a review see~\cite{Ibanez:2007pf}). For this class of models it is predicted that the upper limit on the Higgs boson mass lies below $130\GeV$~\cite{Akula:2011aa, Baer:2011ab, Arbey:2011ab, Draper:2011aa, Carena:2011aa, Akula:2012kk, Arbey:2012dq, Strege:2012bt, Buchmueller:2011ab, Baer:2012mv}. While the general framework of supergravity grand unification allows for non-universalites in the scalar boson sector, the gaugino sector, and the sector with trilinear scalar couplings, we will focus exclusively on the universal case, i.e. mSUGRA. The parameter space of SUGRA models has been analyzed in a number of previous works~\cite{Baer:2012mv, Altunkaynak:2010we, Feldman:2007zn, Feldman:2007fq, Feldman:2008hs, Chen:2010kq, Berger:2008cq, Conley:2010du, Roszkowski:2009ye, Fowlie:2012im, Kim:2013uxa}. Here we will  constrain the parameter space 
so that the Higgs mass in the model is consistent with the LHC data and the relic density is consistent with the WMAP and Planck data. The resulting parameter space consistent with these constraints requires that the light stops lie is a narrow corridor between the LSP mass (which in all cases in this analysis is a neutralino) and 1.2 times the LSP mass.
 
In this region the relic density is satisfied by stop-neutralino coannihilatio {~\cite{ Chen:2010kq, Ellis:2001nx, Santoso:2002xu, Ajaib:2011hs, Raza:2014upa, Yu:2012kj, Harz:2012fz,Ellis:2014ipa,Harz:2014tma,Buchmueller:2015uqa,Bagnaschi:2015eha}}. For this class of models, the mass difference $\Delta m_{\tilde{t}_1,\tilde{\chi}^0_1}$ between the stop and LSP is small and the decay products of the stop are relatively soft. It is found that the lightest stop can have a mass as low as $375\,{\rm GeV}$ and stop pair production cross-sections can be as large as 2~pb. Further, using optimized signal regions it is found that the 375~GeV stop can have a $5\sigma$ discovery
with 60~fb$^{-1}$ of integrated luminosity. Additionally it is found that  in this model stop masses as large as 600 GeV or larger with only a $\sim 40\,{\rm GeV}$ mass gap between the stop and the neutralino can be discovered 
with the design parameters of the LHC RUN-II.

The outline of the rest of the paper is as follows: In section~\ref{sec:spectra} we discuss the parameter space of interest, examining the stop-neutralino coannihilation region of mSUGRA parameter space under the constraints of the Higgs boson mass, WMAP and Planck relic density, and LHC RUN-I exclusion plots on the sparticle mass spectra. In section~\ref{sec:lhc} we carry out a signature analysis of a representative set of parameter points for the LHC RUN-II energy of 14~TeV. Here we analyze various decay channels and signal regions to determine the best avenues for the discovery of this class of models and determine the minimum luminosity needed for the discovery of each of the parameter points discussed. In section~\ref{sec:dm} we investigate the direct detection of  dark matter in models of the type discussed in section~\ref{sec:lhc}. Conclusions are given in section~\ref{sec:conclusion}. 

\section{Parameter Space for Light Stops in mSUGRA}\label{sec:spectra}
An analysis of the parameter space of  mSUGRA was carried out to identify regions with light stops that also satisfy the Higgs boson constraint, the relic density constraint, and the sparticle mass lower limits from LHC RUN-I~\cite{Aad:2015pfx}. The imposition of these constraints drastically reduces the parameter space of models in a manner discussed further below. The sparticle spectrum was generated using \code{SoftSusy 3.5.1}~\cite{Allanach:2001kg} with the mSUGRA input parameter set
\begin{align}
    m_0,~A_0,~m_{\frac{1}{2}},~\tan\beta,~\text{sgn}(\mu)
    \label{eq:params}
\end{align}
where $m_0$ is the universal scalar mass, $m_{1/2}$ is the universal gaugino mass, and $A_0$ is the universal trilinear scalar coupling (all at the grand unification scale); $\tan\beta=\langle H_2 \rangle/\langle H_1\rangle$, where $H_2$ gives mass to the up quarks and $H_1$ gives mass to the down quarks and the leptons, and sgn$(\mu)$ is the sign of the Higgs mixing parameter which enters in the superpotential in the term $\mu H_1 H_2$. The analysis of the relic density was done using \code{Micromegas 4.1.7}~\cite{Belanger:2004yn} and SLHA-format data files were processed using \code{PySLHA}~\cite{Buckley:2013jua}.

\FloatBarrier  
\subsection{Stop-Neutralino Coannihilation Region}\label{sec:2.1}  
\begin{figure}
	\includegraphics[width=0.5\textwidth]{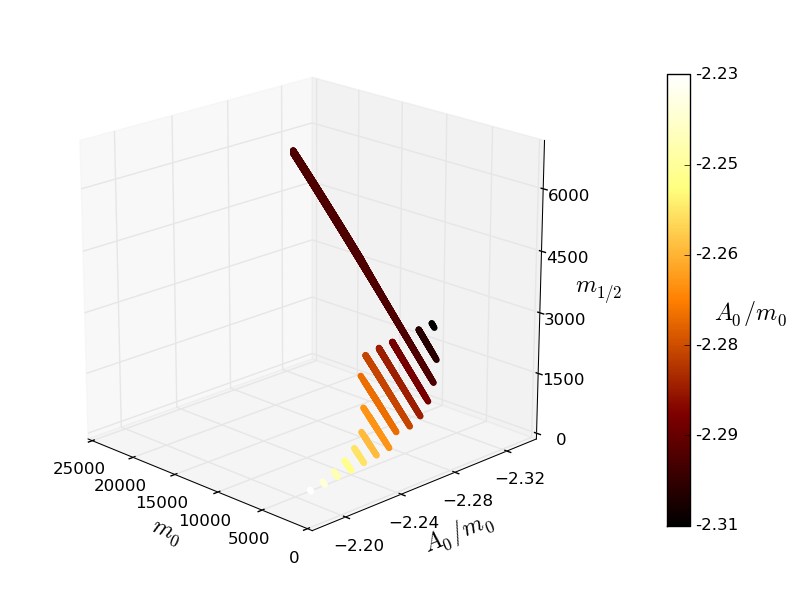}
    \includegraphics[width=0.5\textwidth]{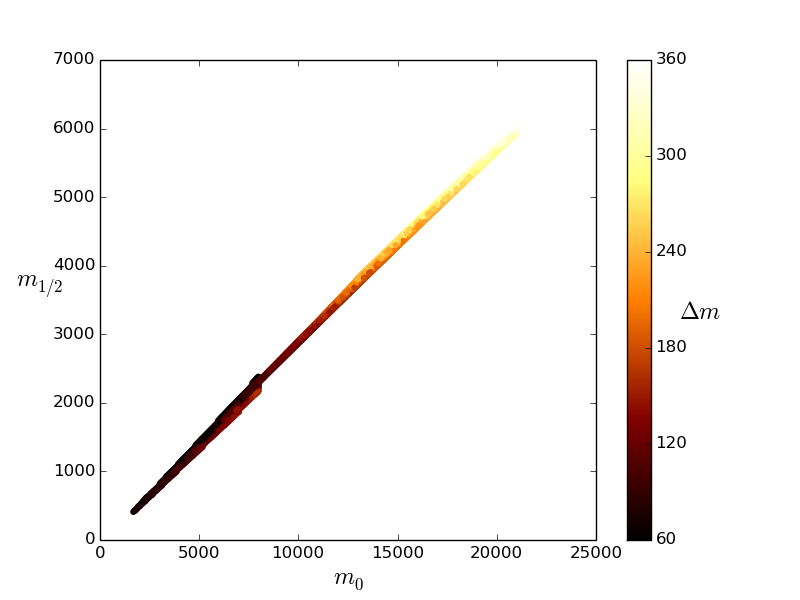}
	\caption{Left panel: A three dimensional plot of the stop-neutralino coannihilation region satisfying eq.~\ref{eq:3} and the Higgs boson mass constraint with three axes chosen as $m_0,~A_0/m_0,$ and $m_{\frac{1}{2}}$ and with $\tan\beta=10$. The image is colored by $A_0/m_0$ values in the range $[-2.2, -2.3]$. Right panel: Two dimensional projection of the left panel in the $m_{\frac{1}{2}}-m_0$ plane. The allowed regions are colored by the stop-neutralino mass gap $\Delta m \equiv m_{\tilde{t}} - m_{\tilde{\chi_1^0}}$.}
	\label{fig:1} 
\end{figure}

Satisfaction of the relic density constraint with a light stop requires stop-LSP coannihilation. The condition of coannihilation is typically
\begin{align}
\frac{\Delta m}{m_{\tilde{\chi}_1^0}} \equiv \frac{{m_{\tilde{t}_1}-m_{\tilde{\chi}_1^0}}}{m_{\tilde{\chi}_1^0}}\leq0.2 
\label{eq:2}
\end{align}
Indeed, eq.~\ref{eq:1} strongly hints that stop-neutralino coannihilation will be the dominant mechanism for satisfying the relic density constraint. It is thus of interest to map out the parameter space where the constraint of eq.~\ref{eq:2} is satisfied. This is exhibited in the two panels of fig.~\ref{fig:1}. The left panel of fig.~\ref{fig:1} gives a three dimensional plot of the coannihilation region with $m_0,~A_0/m_0,$ and $m_{1/2}$ as the three coordinates. The vertical axis on the right gives the color coding for $A_0/m_0$. The analysis shows that the coannihilation region stretches out quite far in $m_0$ and $m_{1/2}$, with $m_0$ extending past 20,000~GeV and $m_{1/2}$ getting as large as 6000~GeV. The section of the curve where $m_0$ and $m_{1/2}$ become very large is part of the hyperbolic branch of radiative breaking of the electroweak symmetry~\cite{Chan:1997bi, Feng:1999mn, Baer:2003wx, Akula:2011jx}. The right panel of fig.~\ref{fig:1} gives the projection of the left panel in the $m_0-m_{1/2}$ plane. The vertical axis on the right gives the color coding for the mass gap between the stop and the neutralino. The analysis of fig.~\ref{fig:1} was done under the Higgs boson mass constraint $126\pm2\GeV$ and the mass gap constraint defined in eq.~3, but no relic density constraints.

It is possible to satisfy the WMAP and Planck relic density constraints in the coannihilation region shown in fig.~\ref{fig:1}. We consider two sets of relic density constraints: (i) the weak relic density constraint $\Omega_{\text{LSP}}<0.12$, and (ii) the strong relic density constraint, which we take to be $0.0946<\Omega_{\text{LSP}}<0.1306$, where the range is $\pm5\sigma$ around the mean WMAP result. The weak relic density constraint allows for multi-component dark matter, i.e. that the relic density is made up of one or more dark matter components other than the neutralino, while the strong relic density constraint  strictly requires that the dark matter is constituted only of the neutralino LSP. The range $\pm5\sigma$ around the mean of WMAP is taken to allow for possible uncertainties in theoretical computations of the relic density. We give now an analysis of the parameter space for these cases.
 
\begin{figure}
\begin{center}
	\includegraphics[width=0.4\textwidth]{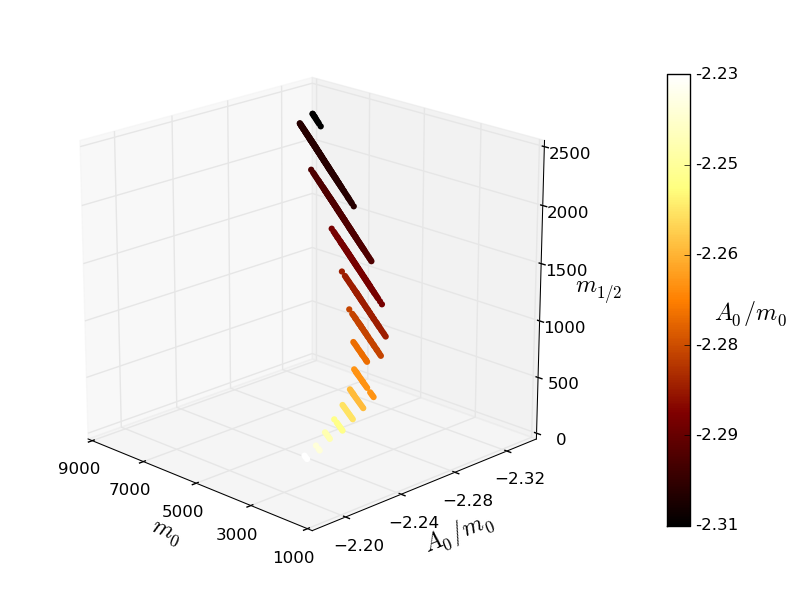}
    \includegraphics[width=0.4\textwidth]{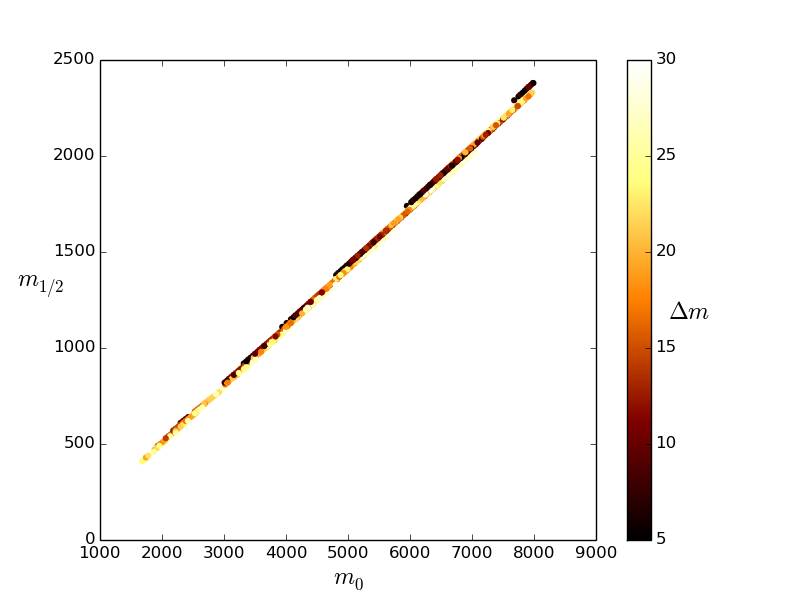}\\
	\includegraphics[width=0.4\textwidth]{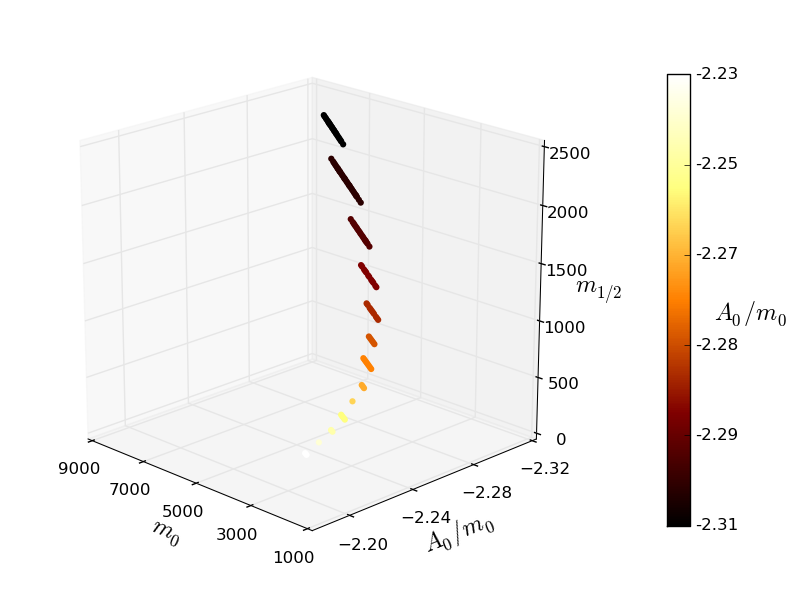}
    \includegraphics[width=0.4\textwidth]{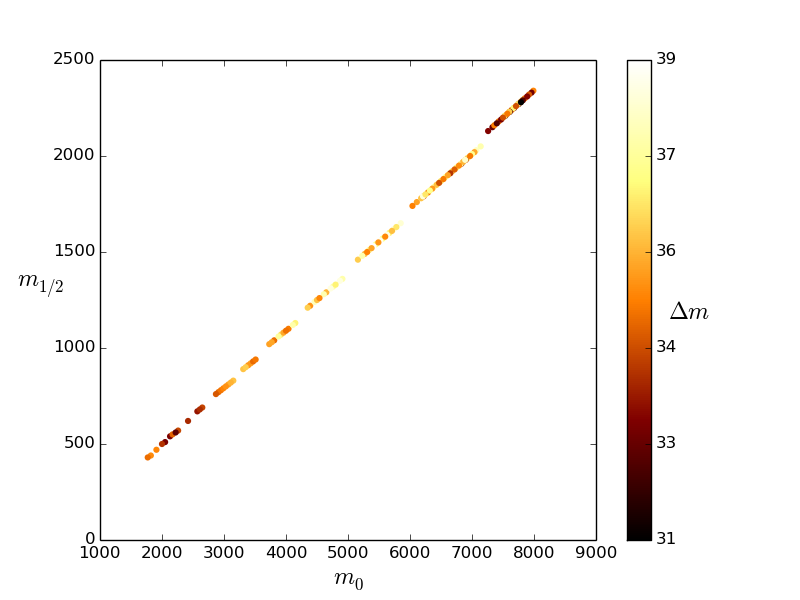}   
    
	\caption{Top panels: Same as the two  panels of fig.~\ref{fig:1} except that relic density constraint $\Omega_{\text{LSP}}<0.12$ is imposed. Bottom panels: Same as the two  panels of fig.~\ref{fig:1} except that the strong relic density constraint $0.0946<\Omega_{\text{LSP}}<0.1306$ is imposed.}
	\label{fig:2} 
	\end{center}
\end{figure}

In the top two panels of fig.~\ref{fig:2}, we investigate the parameter space for the case where we impose the weak relic density constraint (i). The axes in fig.~\ref{fig:2} are  as in fig.~\ref{fig:1}. Here one finds that the major difference between fig.~\ref{fig:1} and the top panels of fig.~\ref{fig:2} is that the coannihilation region has shrunk and the mass gap between the stop and the neutralino is reduced to $\leq45\GeV$. In the bottom two panels of fig.~\ref{fig:2}, we investigate the parameter space for the case where the stronger relic density constraint (ii) is imposed. The two-sided constraint in this case narrows the allowed range of the mass gap between the stop and the neutralino to between 30 and 40\GeV, as exhibited by the vertical bar to the right of the bottom right panel. As in fig.~\ref{fig:1}, the analysis of all panels in fig.~\ref{fig:2} is done under the Higgs boson mass constraint. Similar plots arise for other $\tan\beta$ values, such as $\tan\beta=30$ and $\tan\beta=50$.
    
\begin{figure}
\begin{center}
	\includegraphics[width=0.4\textwidth]{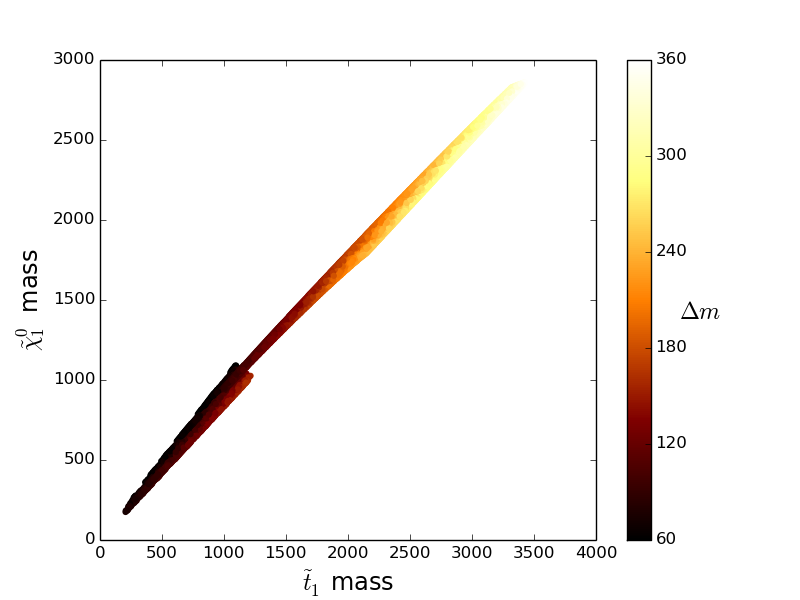}
    \includegraphics[width=0.4\textwidth]{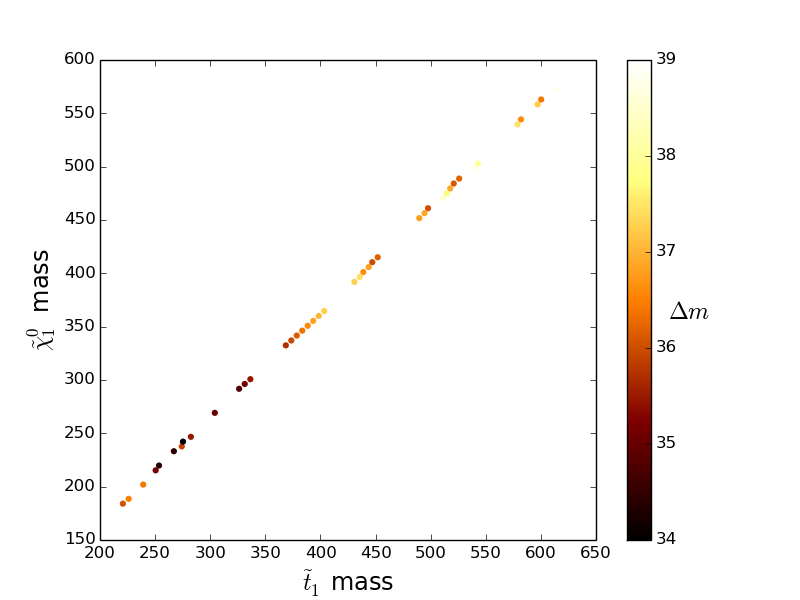}
    \caption{Left panel: A plot of the neutralino mass versus stop mass for the full coannihilation region defined by eq.~\ref{eq:2} and displayed in fig.~\ref{fig:1}. The image is colored by the mass gap $\Delta m$ between the stop and neutralino masses. All masses are in\GeV. Right panel: Same as left panel but restricted to only those points which satisfy the strict relic density constraint (ii) and which are discoverable with a luminosity of $\mathcal{L}\leq3000\,\text{fb}^{-1}$, according to the analysis in section~\ref{sec:lhc}.}
	\label{fig:masses} 
	\end{center}
\end{figure}

The analysis of figs.~\ref{fig:1} and~\ref{fig:2} allows for a few conclusions. While fig.~\ref{fig:1} suggests that the coannihilation region as defined by eq.~\ref{eq:2} continues in $m_0$ past $\sim 25$ TeV, there is an upper cutoff around $m_0\sim 8000$ GeV once the relic density constraint is taken into account, as shown in fig.~\ref{fig:2}. Figure~\ref{fig:masses} displays the $\tilde{t}_1-\tilde{\chi}_1^0$ mass plane and demonstrates that once the relic density constraint is applied, the allowed mass gap is greatly constrained to be between 30 and 40\GeV. From this analysis one concludes that once relic density is considered, the mass gap $\Delta m_{\tilde{t}_1,\tilde{\chi}^0_1}$ always lies below $m_t$ so that the on-shell decay $\tilde{t}_1\to t\tilde{\chi}_1^0$ does not occur, and the dominant decay for the stop in the coannihilation region is $\tilde{t}_1\to c\tilde{\chi}^0_1$. This decay remains dominant in the region where $\Delta m_{\tilde{t}_1,\tilde{\chi}^0_1}$ has an upper limit of $m_W+m_b\sim85\GeV$. Of course the off-shell decay $\tilde{t}_1\to  W^*b\tilde{\chi}_1^0$ can still occur, which will produce signatures of the type $\tilde{t}_1\to bff'\tilde{\chi}^0_1$, where $ff'$ arises from the decay of the $W^*$. The jets arising from the decay will be soft, but high $P_T$ jets could arise from initial and final state radiation (ISR and FSR). 

\section{Signal Analyses for Light Stop Models in LHC RUN-II}\label{sec:lhc}

\begin{figure}[t]
	\includegraphics[width=0.5\textwidth]{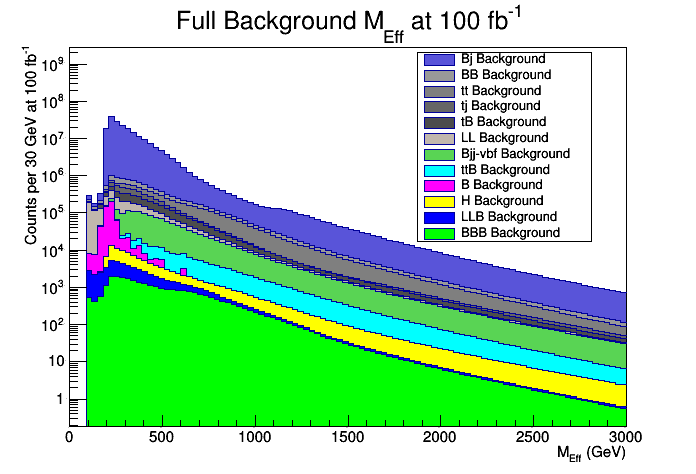}
    \includegraphics[width=0.5\textwidth]{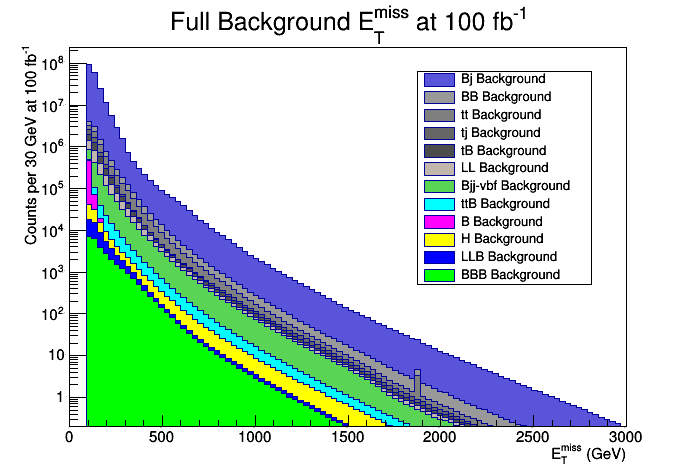}
    \\
	\includegraphics[width=0.5\textwidth]{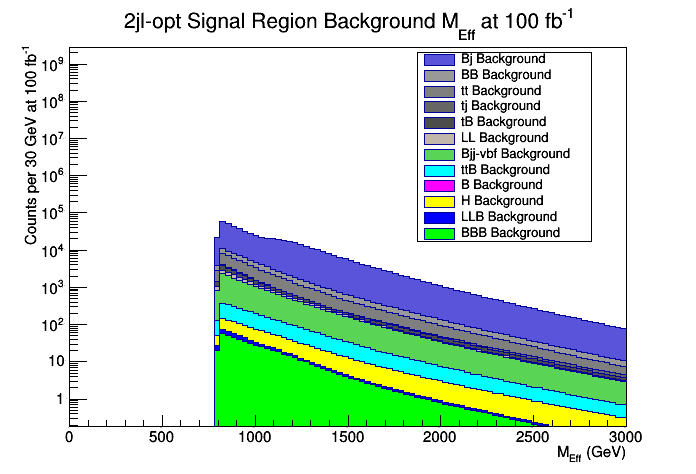}
    \includegraphics[width=0.5\textwidth]{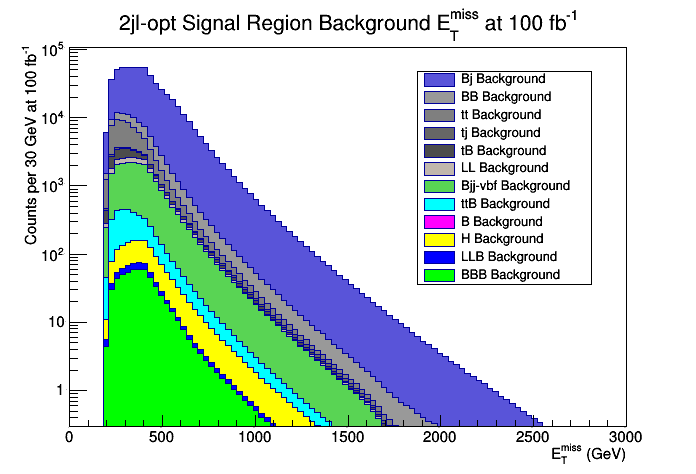}
    \caption{Top panels: Full SNOWMASS SM background after triggering cuts and a pre-cut of $E^{\rm miss}_T\geq 100\,{\rm GeV}$, broken into individual processes and scaled to 100 fb$^{-1}$. Left panel gives  $M_{\rm Eff}({\rm incl.})$ and right panel gives $E^{\rm miss}_T$. Individual data sets are labeled according to eq.~(\ref{Snowlabels}). Bottom panels: $M_{\rm Eff}({\rm incl.})$ (left) and $E^{\rm miss}_T$ (right) for the full SNOWMASS SM background, scaled to 100~fb$^{-1}$, after cuts which define the 2jl-opt signal region (see table~\ref{tab:3} and section~\ref{sec:optim}) are applied.}
	\label{fig:bgs} 
\end{figure}

To assess the appropriate channel for detecting a supersymmetric signal arising from the coannihilation region of parameter space allowed by the Higgs boson mass constraint and the WMAP and Planck relic density constraints, a Monte Carlo analysis was performed for a representative subset of coannihilation region points. The analysis was conducted using \code{MadGraph 2.2.2}~\cite{Stelzer:1994ta, Alwall:2007st, Alwall:2011uj}, which includes \code{Pythia 6.4.28}~\cite{Sjostrand:2006za} for hadronization of decays and \code{Delphes 3.1.2}~\cite{deFavereau:2013fsa} for detector simulation and event reconstruction. \code{MadGraph} calculates the possible decay diagrams for a requested set of decays. From there, the \code{MadEvent} submodule of \code{MadGraph} takes as inputs the sparticle mass spectrum, decay widths, and branching ratios, and uses this information to find the decay cross sections of the chosen decays. With these cross sections \code{MadEvent} can simulate decay events, using \code{Pythia} to complete the hadronization of resulting color-charged particles.  Once event generation is complete, \code{Delphes} is used to reconstruct events within the geometry and triggering environment of a particular detector, outputting a file formatted for later analysis with \code{ROOT 5.34.21}~\cite{Brun:1997}.

For each simulated point of mSUGRA space, the sparticle spectrum was calculated with \code{SoftSusy} and input to \code{SUSY-HIT 1.5}~\cite{Djouadi:2006bz}, which uses \code{SDECAY}~\cite{Muhlleitner:2004mka, Muhlleitner:2003vg} and \code{HDECAY}~\cite{Djouadi:1997yw,Spira:1996if} to calculate the decay widths and branching ratios of the sparticles from the sparticle spectrum. Particular to the coannihilation region spectra studied here, the flavor-violating stop decay module of \code{SUSY-HIT}~\cite{Grober:2014aha} is used in this analysis to obtain more accurate results for the compressed spectrum stop decays. The results calculated by \code{SUSY-HIT} in flavor-violating stop decay mode are inserted back into the general spectrum output given by the vanilla \code{SUSY-HIT} code.

To ensure that every possible decay is considered, Feynman diagrams for the decays p p $\to$ SUSY SUSY were generated in \code{MadGraph}, where ``SUSY'' represents any MSSM particle. Both ISR and FSR are included in the analysis. These diagrams were then used to generate 50,000 MSSM decay events for a collection of points in the coannihilation region of mSUGRA space (see table~\ref{tab:1}), using each point's calculated sparticle mass spectrum, decay widths, and branching ratios, as described above. For each mSUGRA parameter point, \code{MadGraph} generated an overall SUSY decay cross section, as well as cross sections for each individual decay mode, and then randomly generated decays according to those cross sections, hadronizing jets using \code{Pythia}.  Once all 50,000 events were generated, \code{Delphes} completed the detector simulation using the ATLAS detector file, packaging the results into a \code{ROOT} output file.

Search analyses were performed on the generated event set for each mSUGRA parameter point. The analyses were written as \code{ROOT} scripts implementing the search region details for previously published searches at 8~TeV. Specifically, the low-multiplicity jets plus missing transverse energy search signal regions from~\cite{Aad:2014wea} and the monojet search signal regions from~\cite{Aad:2014nra} were implemented and performed on the simulated MSSM decays. 

To compare to the background, the analyses were also performed on the pre-generated SNOWMASS published backgrounds~\cite{Avetisyan:2013onh}. From the total MSSM cross section calculated by \code{MadEvent} and the number of events simulated, an implied luminosity for each point was calculated to allow direct comparison to the backgrounds, which were also scaled individually by implied luminosity.  The many background processes which comprise the SNOWMASS background set are scaled and combined in fig.~\ref{fig:bgs}.  Each individual process is represented by a different color in these plots.  The various background samples are grouped according to the generated final state, with a collective notation given by
\begin{eqnarray} 
J &=& \left\lbrace u,\bar{u} ,d,\bar{d},s,\bar{s} ,c,\bar{c},b,\bar{b} \right\rbrace \, ,\nonumber \\
L &=& \left\lbrace e^+,e^−,\mu^+,\mu^−,\tau^+,\tau^−,\nu_e,\nu_{\mu}, \nu_{\tau}\right\rbrace\, , \nonumber \\
V &=& \left\lbrace W^+,W^−, Z , \gamma \right\rbrace\, , \label{Snowlabels} \\
T &=& \left\lbrace t,\bar{t}\, \right\rbrace \, ,\nonumber \\
H &=& \left\lbrace h_0 \right\rbrace\, . \nonumber 
\end{eqnarray}
in general, events with gauge bosons and SM Higgs bosons in the final state are grouped into a single ``boson'' (B) category. Thus, for example,  the data set ``Bjj-vbf'' represents production via vector boson fusion of a gauge boson or a Higgs with at least two additional light-quark jets.

Signal region cuts are applied to these backgrounds in addition to SUSY Model signals, and then the backgrounds for each process are summed before calculation of the integrated luminosity for discovery. The top panels of fig.~\ref{fig:bgs} illustrate two key kinematic quantities, $M_{\rm Eff}$ and $E_T^{\rm miss}$ for the full background dataset, after minimal cuts for trigger simulation and a $E_T^{\rm miss} \geq 100\,{\rm GeV}$ pre-cut.  Here and throughout,  $M_{\rm Eff}$ is specifically  $M_{\rm Eff}(\rm incl.)$, defined as the scalar sum of $E_T^{\rm miss}$ and the $p_T$ of all jets with $p_T(j) \geq 40\,{\rm GeV}$.
The lower panels demonstrate the same quantities in the full background after applying the specific cuts which define signal region 2jl-opt, defined later in the text (see section~\ref{sec:optim}).

\subsection{LHC Production and Signal Definitions}
In table~\ref{tab:1} we present a set of model points in the mSUGRA parameter space with light stops which give a Higgs boson mass consistent with LHC measurements and have a relic density consistent with WMAP and Planck observations. For the sake of consistency, all points in table~\ref{tab:1} assume $\tan\beta=10$, though the results we will discuss are largely insensitive to the precise value of this parameter. Our simulation of these points at $\sqrt s= 8$ TeV shows that they would have escaped detection using the total integrated luminosity accumulated at LHC RUN-I~\cite{Aad:2015pfx}. In all cases, for the parameter points listed in table~\ref{tab:1} the light stop is the NLSP and its dominant decay mode is to a light chargino and a charm quark, which is a flavor-changing process. The other lightest particles are the second lightest neutralino $\tilde{\chi}_2^0$, the light chargino $\tilde{\chi}_1^{\pm}$, and the gluino $\tilde{g}$. Thus in the model points of table~\ref{tab:1}, the sparticle mass hierarchy is\footnote{The hierarchy eq.~\ref{eq:3} is mSP[t1a] in the notation of Table 5 of \cite{Francescone:2014pza}.}
\begin{align}
    m_{\tilde{\chi}_1^0}<m_{\tilde{t}_1}<m_{\tilde{\chi}_2^0}<m_{\tilde{\chi}_1^{\pm}}<m_{\tilde{g}}\, .
\label{eq:3}
\end{align}
Each of the NNLSP decays, i.e. the decays of $\tilde{\chi_2}^0,~\tilde{\chi}_1^{\pm},$ and $\tilde{g}$, involves a light stop in their dominant decay. However, the decay of $\tilde{t}_1$ in the coannihilation region will yield soft jets because of the small mass gap between its mass and the LSP mass, which makes it difficult to observe it directly.

\begin{table}[t]
	\centering
	\begin{tabulary}{\linewidth}{l|ccc|ccccccc}
	Model & $m_0$ & $A_0$ & $m_{\frac{1}{2}}$ & $m_{\tilde{t}_1}$ & $m_{\tilde{\chi}^0_1}$  & {$\Delta m_{\tilde{t}_1,\tilde{\chi}^0_1}$} & $m_h$ & $m_{\tilde{t}_2}$ & $m_{\tilde{g}}$ & $m_{\tilde{\chi}^0_2} \simeq m_{\tilde{\chi}^\pm_1}$\\
	\hline
    i.    & 2910 & -6547 & 770 & 374 & 337 & {37} & 124.5 & 2316 &  1865 & 652 \\
    ii.   & 2990 & -6727 & 790 & 383 & 346 & {37} & 125.6 & 2376 &  1900 & 669 \\
    iii.  & 3150 & -7087 & 830 & 403 & 365 & {38} & 127.9 & 2497 &  1989 & 704 \\
    iv.   & 3350 & -7570 & 900 & 436 & 397 & {39} & 124.8 & 2663 &  2142 & 764 \\
    v.    & 3470 & -7842 & 930 & 447 & 410 & {37} & 127.1 & 2754 &  2208 & 790 \\
    vi.   & 3730 & -8467 & 1020 & 490 & 452 & {38} & 124.1 & 2968 &  2403 & 868 \\
    vii.  & 3810 & -8648 & 1040 & 498 & 461 & {37} & 125.3 & 3029 &  2447 & 885 \\
    viii. & 3920 & -8898 & 1070 & 515 & 475 & {40} & 125.5 & 3114 &  2512 & 911 \\
    ix.   & 4150 & -9420 & 1130 & 543 & 503 & {40} & 127.8 & 3289 &  2644 & 964 \\
    x.    & 4350 & -9918 & 1210 & 579 & 540 & {39} & 124.1 & 3462 &  2813 & 1032 \\
    xi.   & 4460 & -10168 & 1240 & 595 & 554 & {41} & 124.4 & 3547 &  2878 & 1059 \\
    xii.  & 4610 & -10510 & 1280 & 614 & 572 & {42} & 125.4 & 3692 & 2965 & 1093 \\
	\hline
	\end{tabulary}
	\caption{The subset of mSUGRA parameter points which satisfy the Higgs boson mass constraint and the relic density constraint in the stop coannihilation region and which were chosen for LHC analysis. The mSUGRA parameters are given (all points have $\tan\beta=10$ and $\mu>0$), followed by key superpartner masses (in\GeV). Our simulations show that the stop-neutralino mass combinations arising from Models~i-xii lie outside of the exclusion plots of RUN-I of the LHC.}
    \label{tab:1}
\end{table}

In table~\ref{tab:2} we give the production processes involving $\tilde{t}_1,~\tilde{\chi}_2^0,~\tilde{\chi}_1^{\pm},$ and $\tilde{g}$ at RUN-II of the LHC for each of the parameter points listed in table~\ref{tab:1}. The dominant production modes generally consist of the following set:
\begin{align}
    gg\to\tilde{t}_1\tilde{t}_1,~qq\to\tilde{t}_1\tilde{t}_1,~qq\to\tilde{\chi}^0_2\tilde{\chi}^+_1,~qq\to\tilde{\chi}^+_1\tilde{\chi}^-_1,~gg\to\tilde{g}\tilde{g},\text{ and }qq\to\tilde{g}\tilde{g} \, .
    \label{eq:production}
\end{align}
The largest production cross section is for the stops, followed by the production of the weak gauginos, and then the gluino. {For the parameter regime presented in table~\ref{tab:1}, the typical mass difference  $\Delta m_{\tilde{t}_1,\tilde{\chi}^0_1}$ is consistently near 40~GeV, with the flavor-violating decay $\tilde{t}_1 \to c \tilde{\chi}^0_1$ {typically} representing 60\% or more of the total decay width of the lightest stop.}

\begin{table}
	\centering
	\begin{tabulary}{\linewidth}{l|*{7}{C}}
	Model & $gg\to\tilde{t}_1\tilde{t}_1$ & $qq\to\tilde{t}_1\tilde{t}_1$ & $qq\to\tilde{\chi}^0_2\tilde{\chi}^+_1$ & $qq\to\tilde{\chi}^+_1\tilde{\chi}^-_1$ & $gg\to\tilde{g}\tilde{g}$ & $qq\to\tilde{g}\tilde{g}$ & BR($\tilde{t}_1\tilde{t}_1\to c\tilde{\chi}_1^0$)\\
	\hline
    i.    & 2.0 & 0.25 & 1.0$\times 10^{-2}$ & 4.8$\times 10^{-3}$ & 9.0$\times 10^{-4}$ & 2.1$\times 10^{-4}$ & 89.7\% \\
    ii.   & 1.7 & 0.22 & 8.8$\times 10^{-3}$ & 4.3$\times 10^{-3}$ & 6.9$\times 10^{-4}$ & 1.8$\times 10^{-4}$ & 88.0\% \\
    iii.  & 1.3 & 0.18 & 6.9$\times 10^{-3}$ & 3.3$\times 10^{-3}$ & 4.1$\times 10^{-4}$ & 1.2$\times 10^{-4}$ & 83.9\% \\
    iv.   & 0.85 & 0.12 & 4.6$\times 10^{-3}$ & 2.1$\times 10^{-3}$ & 1.8$\times 10^{-4}$ & 5.9$\times 10^{-5}$ & 81.4\% \\
    v.    & 0.74 & 0.11 & 3.9$\times 10^{-3}$ & 1.8$\times 10^{-3}$ & 1.3$\times 10^{-4}$ & 4.4$\times 10^{-5}$ & 86.0\% \\
    vi.   & 0.44 & 7.1$\times 10^{-2}$ & 2.4$\times 10^{-3}$ & 1.1$\times 10^{-3}$ & 3.8$\times 10^{-5}$ & 1.8$\times 10^{-5}$ & 80.9\% \\
    vii.  & 0.40 & 6.6$\times 10^{-2}$ & 2.4$\times 10^{-3}$ & 1.0$\times 10^{-3}$ & 3.2$\times 10^{-5}$ & 1.5$\times 10^{-5}$ & 83.4\%\\
    viii. & 0.33 & 5.6$\times 10^{-2}$ & 1.9$\times 10^{-3}$ & 8.9$\times 10^{-4}$ & 2.1$\times 10^{-5}$ & 1.1$\times 10^{-5}$ & 74.7\%\\
    ix.   & 0.24 & 4.3$\times 10^{-2}$ & 1.4$\times 10^{-3}$ & 6.4$\times 10^{-4}$ & 9.4$\times 10^{-6}$ & 6.3$\times 10^{-6}$ & 71.0\%\\
    x.    & 0.17 & 3.1$\times 10^{-2}$ & 9.1$\times 10^{-4}$ & 4.3$\times 10^{-4}$ & 3.9$\times 10^{-6}$ & 3.0$\times 10^{-6}$ & 71.9\%\\
    xi.   & 0.14 & 2.7$\times 10^{-2}$ & 7.9$\times 10^{-4}$ & 3.7$\times 10^{-4}$ & 2.9$\times 10^{-6}$ & 2.2$\times 10^{-6}$ & 61.5\%\\
    xii.  & 0.12 & 2.3$\times 10^{-2}$ & 6.6$\times 10^{-4}$ & 3.0$\times 10^{-4}$ & 1.6$\times 10^{-6}$ & 1.5$\times 10^{-6}$ & 60.9\%\\
	\hline
	\end{tabulary}
	\caption{ Production cross sections in pb for LHC RUN-II of the dominant supersymmetric modes $gg\to\tilde{t}_1\tilde{t}_1$, $qq\to\tilde{t}_1\tilde{t}_1$, $qq\to\tilde{\chi}^0_2\tilde{\chi}^+_1$, $qq\to\tilde{\chi}^+_1\tilde{\chi}^-_1$, $gg\to\tilde{g}\tilde{g}$, and $qq\to\tilde{g}\tilde{g}$, as well as the branching ratio for the process $\tilde{t}_1\tilde{t}_1\to c\tilde{\chi}_1^0$, for the set of parameter points given in table~\ref{tab:1}.}
	\label{tab:2}
\end{table}

Given the dominance of stop production followed by $\tilde{t}_1 \to c \tilde{\chi}^0_1$, our signature analysis concentrates on two ATLAS searches: the low-multiplicity jets plus missing transverse energy $E_T^{\rm miss}$ search~\cite{Aad:2014wea}, and the dedicated stop search of~\cite{Aad:2014nra}.\footnote{An alternative set of signal regions for supersymmetry discovery are used by the CMS Collaboration, see e.g., \cite{Khachatryan:2015pwa,Khachatryan:2015exa} which involves razor variables.}
The multi-jet search defines fourteen signal regions with jet multiplicities between 2~and~6 jets, and with varying requirements on the inclusive effective mass $M_{\rm Eff}$(inc.), designated as ``loose'' (low $M_{\rm Eff}$(inc.)) to ``tight'' (high $M_{\rm Eff}$(inc.)). The dominant signal for the points in table~\ref{tab:1} involves typically 2-4 reconstructed jets with $p_T(j) \geq 20\,{\rm GeV}$ and a low effective mass. Examination of all 14~signatures therefore reveals that the two-jet ``loose'' signature (2jl) is the most effective topology for searching for these models. This signal region has the  selection requirements as listed in {the left panel of} table~\ref{tab:3}.

\begin{table}
	\centering
	\begin{tabulary}{0.5\linewidth}{l|l}
	Requirement {(2jl SR)} & Value \\
    \hline
    $E^{\text{miss}}_T\text{ (GeV)}$ &  $>$160 \\
    $p_T(j_1)\text{ (GeV)}$ & $>$130 \\
    $p_T(j_2)\text{ (GeV)}$ & $>$ 60 \\
    $\Delta\phi(\text{jet}_{1,2},E^{\text{miss}}_T)_{\text{min}}$ & $>$ 0.4 \\
    $E^{\text{miss}}_T/\sqrt{H_T}(\text{ (GeV}^{1/2})$ & $>$ 8 \\
    $M_{\text{Eff}}(\text{inc.})\text{ (GeV)}$ & $>$ 800 \\
	\hline
	\end{tabulary}
    \quad
    \begin{tabulary}{0.5\linewidth}{L|l}
    Requirement {(M1 SR)} & Value \\
    \hline
    $E^{\text{miss}}_T\text{ (GeV)}$ & $>$ 220 \\
    $p_T(j_1)\text{ (GeV)}$ & $>$ 280 \\
    $\Delta\phi(\text{jet},E^{\text{miss}}_T)_{\text{min}}$ &  $>$ 0.4 \\
    At most 2 other jets with $p_T\text{ (GeV)}$ & $>$ 30 \\
    \end{tabulary}
	\caption{{Left:} The selection criteria used for the signal region 2jl, where ``2j'' stands for two jets and ``l'' stands for loose in the nomenclature of Table 2 of the ATLAS analysis~\cite{Aad:2014wea}. This signal region is found to be the most dominant for the parameter set given in table~\ref{tab:1}. {Right: The selection criteria for the monojet signal region M1 from the ATLAS dedicated stop search~\cite{Aad:2014nra}. This signal region is subdominant for the parameter set of interest.}}
	\label{tab:3}
\end{table}

{The dedicated stop search of~\cite{Aad:2014nra} involves two types of topology. The first is a monojet-like signature, divided into three signal regions---M1, M2, and M3---defined by increasingly stringent requirements on the $p_T(j_1)$ of the leading jet and the missing transverse energy. The second topology also involves a hard leading jet, with  $p_T(j_1)$ requirements similar to that of the monojet analysis, but also requires a charm tag on at least one of the sub-leading jets. The multi-variate technique that generates the charm tag is not something that is easily reproduced, nor modeled in {\tt Delphes}. We therefore focus only on the monojet searches in this work.} {The most successful monojet search for this parameter region is M1, which has the selection requirements specified in the right panel of table~\ref{tab:3}.}

\begin{table}
	\centering
	\begin{tabulary}{0.5\linewidth}{l|CCC}
	Model & {$\mathcal{L}$ for $5\sigma$ discovery in 2jl-opt SR} & $\mathcal{L}$ for $5\sigma$ discovery in 2jl {SR} & $\mathcal{L}$ for $5\sigma$ discovery in M1 SR\\
	\hline
    i.    & 61 & 88 & 369\\
    ii.   & 79 & 112& 396\\
    iii.  & 100 & 135 & 535\\
    iv.   & 148 & 202 & 739\\
    v.    & 180 & 242 & 777\\
    vi.   & 322 & 446 & 1561\\
    vii.  & 367 & 522 & 1881\\
    viii. & 502 & 706 & 2460\\
    ix.   & 725 & 1025 & 4035\\
    x.    & 1214 & 1716 & 6118\\
    xi.   & 1621 & 2248 & 8539\\
    xii.  & 2140 & 2994 & 10538\\
	\hline
	\end{tabulary}
	\caption{Analysis of the discovery potential for supersymmetry for the parameter space of table~\ref{tab:1}, using the leading channel 2jl, the sub leading signal channels M1, {and the optimized channel 2jl-opt,} where the minimum integrated luminosity needed for $5\sigma$ discovery is given in $\text{fb}^{-1}$.}
	\label{tab:4}
\end{table}

Using the techniques described above, we analyzed each of the model points of table~\ref{tab:1} to identify a minimum required luminosity for $5\sigma$ {$\frac{\text{signal}}{\sqrt{\text{background}}}$} discovery in the signal regions {of interest. { In the analysis here, only statistical errors were considered and no attempt was made to estimate systematic errors.   Such an effort requires detailed and specialized knowledge of the experimental apparatus.  Furthermore, systematic errors at $\sqrt{s}=14$ TeV are expected to be very different in form and magnitude than the published systematic errors at $\sqrt{s}=8$~TeV and at lower luminosity.
Our estimates of the integrated luminosities given  here are based on single-channel analyses, and
one expects that combining channels can potentially result in discovery with even smaller integrated luminosities, even after accounting for systematic error.}\\

In addition, because the tightly constrained parameter space yields a similarly constrained sparticle spectrum and signal, variations of these signal regions were tested in order to optimize them for sparticle spectra with the particular properties of tables~\ref{tab:1} and~\ref{tab:2} (see section~\ref{sec:optim}).} The {minimum integrated luminosity} results of these analyses are presented in table~\ref{tab:4}. It is found that these model points will be discoverable with integrated luminosities beginning at {$\sim$90 fb$^{-1}$ using existing searches or $\sim$60 fb$^{-1}$ using an optimized search, described in section~\ref{sec:optim}.

\begin{figure}
\begin{center}
	\includegraphics[width=0.48\textwidth]{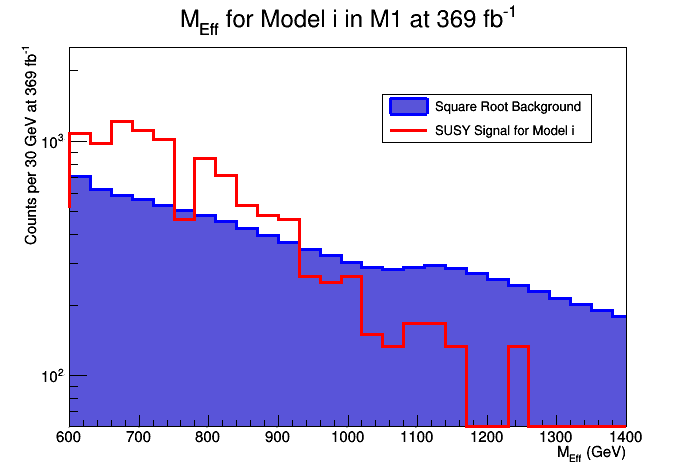}
    \includegraphics[width=0.48\textwidth]{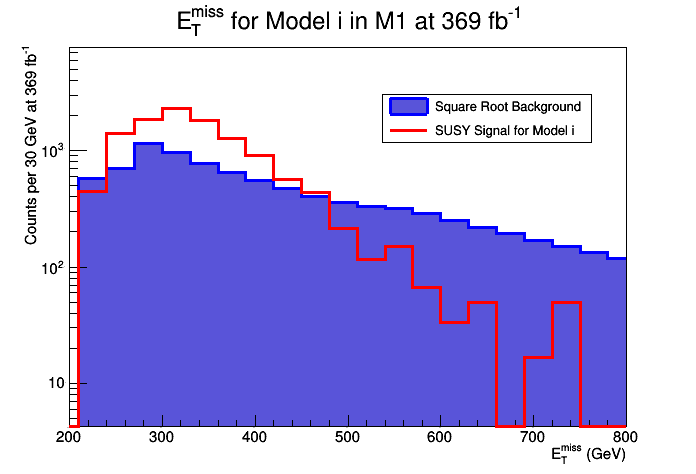}
    \caption{Left panel: Distribution in $M_{\text{Eff}}$ for the M1 signal region for Model~i. Plotted is the number of counts for the SUSY signal and the square root of the total SM backgrounds. The analysis is done at 369 fb$^{-1}$ of integrated luminosity, which gives a $5\sigma$ discovery in this signal region. Right panel: The same analysis as in the left panel but for $E^{\text{miss}}_T$.}
	\label{fig:histM1} 
	\end{center}
\end{figure}

\begin{figure}
\begin{center}
	\includegraphics[width=0.48\textwidth]{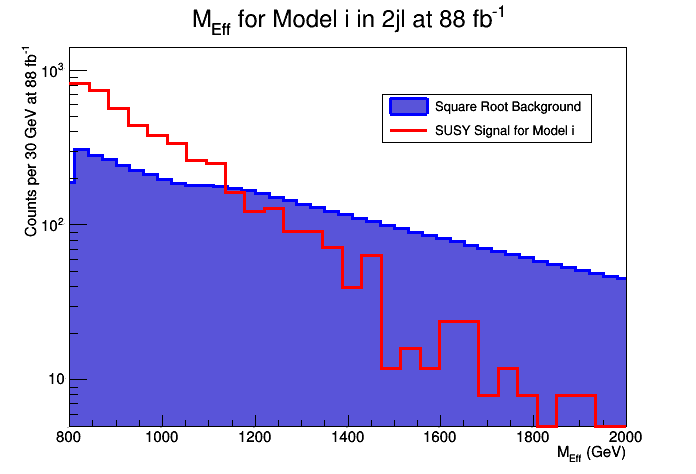}
    \includegraphics[width=0.48\textwidth]{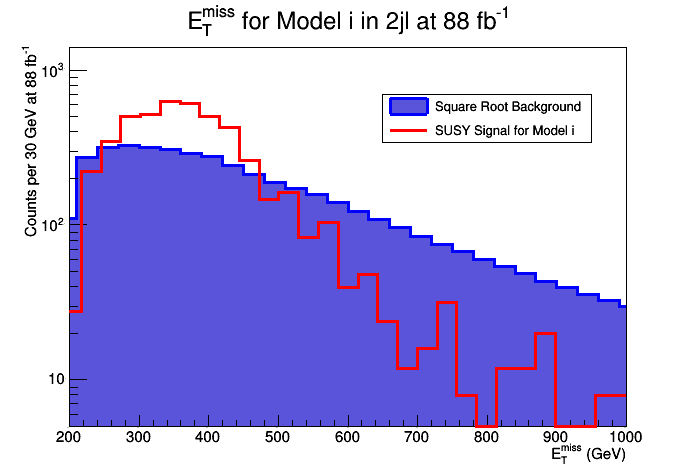}
    \caption{ Left panel: Distribution in $M_{\text{Eff}}$ for the 2jl signal region for Model~i. Plotted is the number of counts for the SUSY signal and the square root of the total SM backgrounds. The analysis is done at 88 fb$^{-1}$ of integrated luminosity, which gives a $5\sigma$ discovery in this signal region. Right panel: The same analysis as in the left panel but for $E^{\text{miss}}_T$.}
	\label{fig:hist2jl} 
	\end{center}
\end{figure}    

Results for the M1 signal region are represented in fig.~\ref{fig:histM1}, which displays the Model~i signal and the square root of the summed background after M1 cuts.  The left panel plots $M_{\text{Eff}}$ while the right panel plots $E^{\text{miss}}_T$. These figures are shown at 369 fb$^{-1}$, the necessary integrated luminosity for $5\sigma$ discovery for Model~i using signal region M1.

Likewise, a representative result for the 2jl signal region is shown in fig.~\ref{fig:hist2jl}.  Like fig.~\ref{fig:histM1}, this displays the signal for Model~i superimposed upon the square root of the total summed background, both after cuts, this time in 2jl.  The left panel plots $M_{\text{Eff}}$ and the right panel plots $E^{\text{miss}}_T$.  The integrated luminosity used is 88 fb$^{-1}$, which gives $5\sigma$ discovery of Model~i using signal region 2jl.

Comparing the results of table~\ref{tab:4} for the dominant signal and for the subdominant signals, we find that the integrated luminosity needed for the subdominant signal {M1} can be larger by a factor of 3 or more. {All other signal regions from \cite{Aad:2014wea} and \cite{Aad:2014nra} that were analyzed required even higher integrated luminosities for $5\sigma$ discovery. This verifies the claim that the 2jl is the dominant signal for discovery of the models listed in table~\ref{tab:1}.  However, because 2jl is a general-purpose search optimized at 8~TeV, it is possible to improve upon its performance for this parameter region.

\subsection{{Optimizing the Signal Regions}}\label{sec:optim}

As mentioned above, because the leading signal region 2jl is a general-purpose signal region developed for application at 8~TeV, it should be possible to improve on its performance by closely examining the cuts made to certain key parameters. Three primary changes were identified for analysis and comparison to the baseline signal regions: Relaxing the kinematic cuts on $M_{\rm Eff}$(inc.), $E^{\text{miss}}_T$, or $p_T(j)$, increasing the cut on $\Delta\phi$ between $E^{\text{miss}}_T$ and the two leading jets, and making an additional cut on a new parameter given by the ratio $p_T(j_1)/E^{\text{miss}}_T$. Optimizations on these parameters were performed on the 2jl and M1 signal regions described in table~\ref{tab:3}, as well as on signal region 3j from \cite{Aad:2014wea}, which differs from 2jl primarily by requiring a third jet with $p_T(j_3)>60\GeV$, and by imposing a much tighter cut $M_{\rm Eff} > 2200\GeV$ (as opposed to $M_{\rm Eff} >800\GeV$).

With values of $\Delta m_{\tilde{t}_1,\tilde{\chi}^0_1}$ of around 40 GeV, there is only energy available in the $\tilde{t}_1\to c\tilde{\chi}^0_1$ to make a fairly soft jet---ISR is generally relied upon to produce high-$p_T$ jets in the final state.  Thus, it is reasonable that relaxing the kinematic cuts on $M_{\rm Eff}$(inc.)~could yield an improved signal region.  For the case of 2jl, reducing the $M_{\rm Eff}$(incl.)~cut from 800~GeV to 400~GeV improved the result from the base case by almost a factor of~2 for the lightest points in the parameter space. But as parameter points become heavier this advantage decreases, and eventually the modified signal region becomes worse.  This trend is reflected also if the kinematic cuts in M1 are reduced so that both $E^{\text{miss}}_T>150\GeV$ and $p_T(j_1)>150\GeV$ are required; performance of the signal region improves for the lightest points but becomes worse for higher-mass regions of the parameter space.  Reducing the $M_{\rm Eff}$(incl.)~cut on 3j from $2200\GeV$ to $800\GeV$, a value that matches the cut for 2jl, dramatically improves the 3j performance, leaving it only slightly worse than 2jl.  In fact, after this change there is little that distinguishes the two signal regions; the modified 3j requires a third jet with a minimum $p_T(j_3)>60\GeV$, while 2jl does not look for a third jet.  Thus, 3j is dropped from consideration because its optimization evolves it into 2jl.

\begin{figure}[t]
\begin{center}
	\includegraphics[width=0.48\textwidth]{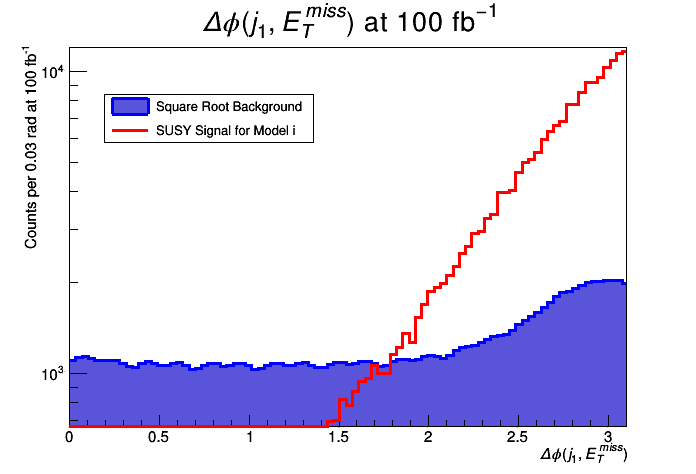}
    \includegraphics[width=0.48\textwidth]{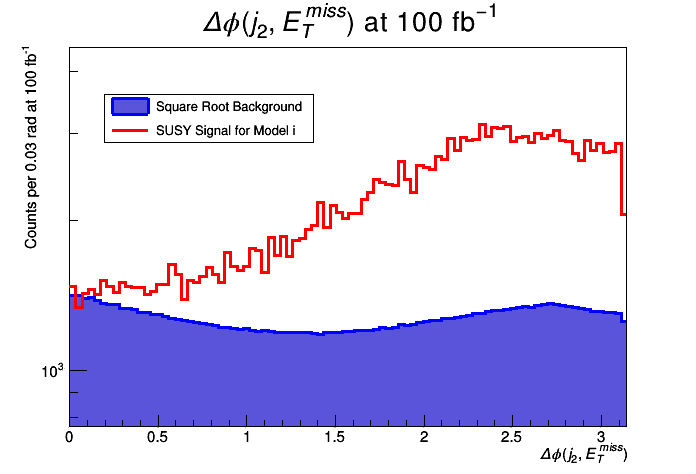}
    \caption{Left panel: Distribution of $\Delta\phi(j_1,E^{\text{miss}}_T)$ against the square root of the full SM background at 100 fb$^{-1}$ for Model~i before any cuts are applied.  Right panel: The same as in the left panel but for $\Delta\phi(j_2,E^{\text{miss}}_T)$}
	\label{fig:deltaphi} 
	\end{center}
\end{figure}

Because the signal is expected to consist primarily of a single jet recoiling against $E^{\text{miss}}_T$, a higher $\Delta\phi(j,E^{\text{miss}}_T)$ is expected in the signal relative to the more uniform background.  This is demonstrated in fig.~\ref{fig:deltaphi}, where the left panel shows the shape of $\Delta\phi(j_1,E^{\text{miss}}_T)$ against the {square root of the summed} background for Model~i before any cuts are applied, and the right panel shows the same for $\Delta\phi(j_2,E^{\text{miss}}_T)$.  Based upon these results, $\Delta\phi$ cuts were increased from 0.4 to $\frac{\pi}{2}$ for $j_1$ and from 0.4 to 1.0 for $j_2$.  Making this change improved the 2jl results by $\sim$30\%, yielding the signal region described in table~\ref{tab:4} as 2jl-opt.  Making this same change in the M1 signal region yielded a poorer search; the integrated luminosity required for $5\sigma$ discovery increased.

Finally, motivated by recent suggestions in~\cite{An:2015uwa,Macaluso:2015wja} ,we considered including an additional kinematic parameter, $r=p_T(j_1)/E^{\text{miss}}_T$. Because this parameter measures the degree to which the recoiling missing energy is concentrated in a single jet, it was expected that it would help distinguish this signal. However, when $r>0.5$ was applied to 2jl and M1, with and without optimization in $\Delta\phi$, the new signal region was worse in every case.

After these investigations, the final optimal signal region was deemed to be 2jl-opt, which, as described above, is the same as 2jl (see table~\ref{tab:3}) but with $\Delta\phi$ cuts increased from 0.4 to $\frac{\pi}{2}$ for $j_1$ and from 0.4 to 1.0 for $j_2$.  This signal region boasts a consistent 30\% increase in performance as compared to 2jl across the full range of the parameter space.

In fig.~\ref{fig:hist3} we exhibit $M_{\text{Eff}}$ {and $E^{\text{miss}}_T$} distributions in the {optimized} signal region {2jl-opt} for three selected models (i, vii, and xii). Plotted are the number of counts per 30~GeV energy bins for the SUSY signal and the square root of the full Standard Model backgrounds. In the left panel of each figure we give the analysis for $M_{\rm Eff}$ and in the right we give the analysis for $E^{\text{miss}}_T$. 
The top panels show the analysis for Model~i. Here the distribution is given at {61} fb$^{-1}$ of integrated luminosity, which is found to be minimum integrated luminosity which gives a $5\sigma$ discovery signal for supersymmetry in {2jl-opt}. Identical analyses are given in the middle panels for Model~vii, but at {367} fb$^{-1}$ of integrated luminosity, which gives a $5\sigma$ discovery signal for that point, and in the bottom panels for Model~xii. Here one finds that a $5\sigma$ discovery requires a minimum of {2140} fb$^{-1}$ of integrated luminosity. Thus Models~i, vii, and~xii are all eventually discoverable with the design parameters of the machine at $\sqrt{s}=14\,{\rm TeV}$.

\begin{figure}
\begin{center}
	\includegraphics[width=0.48\textwidth]{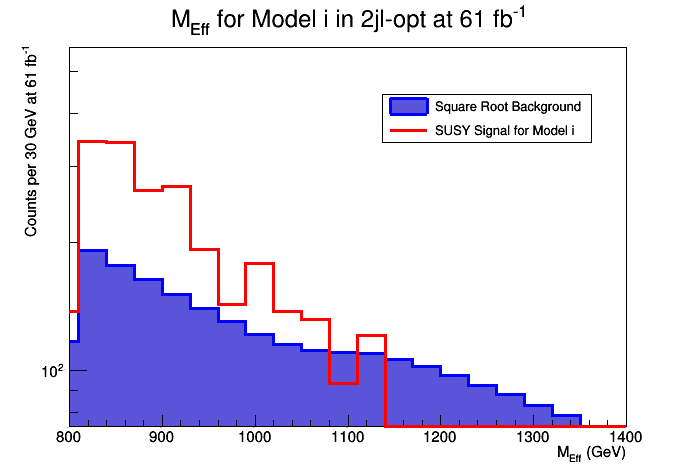}
    \includegraphics[width=0.48\textwidth]{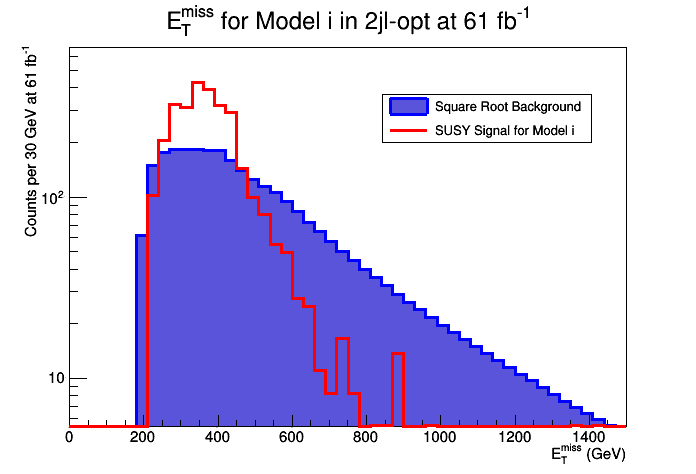}
\\
	\includegraphics[width=0.48\textwidth]{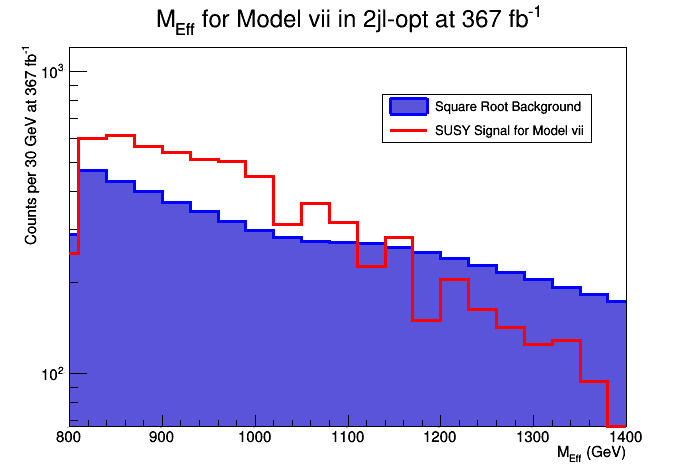}
    \includegraphics[width=0.48\textwidth]{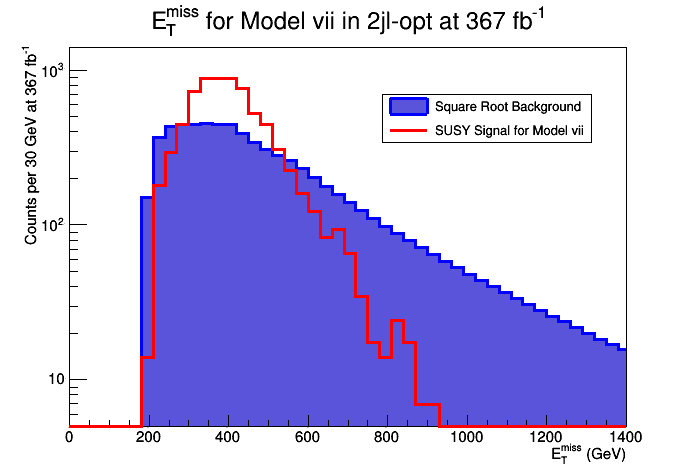}
\\
	\includegraphics[width=0.48\textwidth]{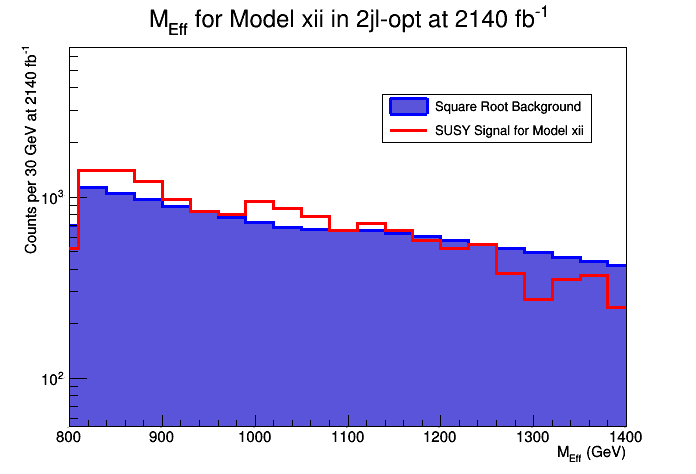}
    \includegraphics[width=0.48\textwidth]{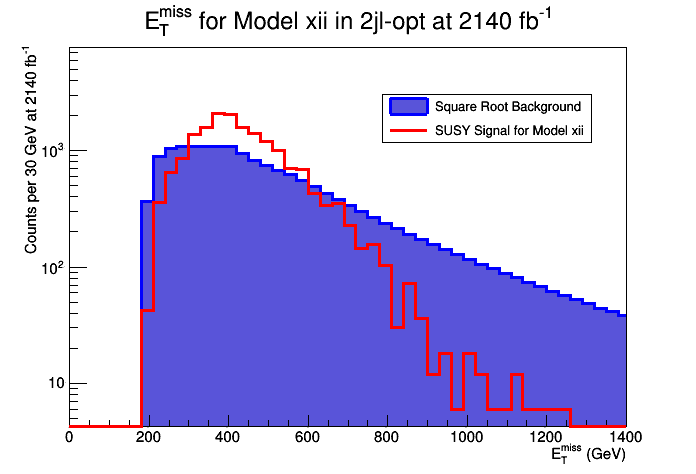}
    \caption{Distribution in $M_{\rm Eff}$ (left panels) and $E_T^{\rm miss}$ (right panels) for the optimized two-jet signal region 2jl-opt. Top panels show Model~i normalized to 61 fb$^{-1}$ of integrated luminosity, while the middle and bottom panels give Models~vii and~xii, normalized to 367 fb$^{-1}$ and 2140 fb$^{-1}$ of integrated luminosity, respectively. These numbers give a $5\sigma$ discovery in this signal region for the three model points. Plotted in all three cases is the number of counts for the SUSY signal and the square root of the total SM background. }
	\label{fig:hist3} 
	\end{center}	
\end{figure}

\section{LHC and Dark Matter}\label{sec:dm}

\begin{table}[t]
	\centering
	\begin{tabulary}{\linewidth}{l|CC}
Model & $\sigma^{SI}_{p\chi_1^0}\times 10^{48}$ & $\sigma^{SD}_{p\chi_1^0}\times 10^{46}$\\
    \hline
    i.    & 3.5 & 9.6\\
    ii.   & 3.2 & 8.6 \\
    iii.  & 2.6 & 6.9 \\
    iv.   & 2.6 & 5.2\\
    v.    & 2.2 & 4.5 \\
    vi.   & 2.1& 3.2\\
    vii.  & 1.9 & 3.0 \\
    viii. & 1.8 & 2.6 \\
    ix.   & 1.5 & 2.1 \\
    x.    & 1.6 & 1.7 \\
    xi.   & 1.5 & 1.5 \\
    xii.  & 1.4 & 1.3 \\
	\hline
	\end{tabulary}
\caption{CDM  proton-neutralino spin-independent ($\sigma^{SI}_{p\chi_1^0}$) 
and spin-dependent ($\sigma^{SD}_{p\chi_1^0}$)
cross sections in units of cm$^{-2}$ for the set of model points of table~\ref{tab:1}.
None of the parameter points are ruled out by the current dark matter experiments, 
in particular by LUX.
}
    \label{tab:8}
\end{table}

Interesting connections between LHC physics and dark matter have been noted in previous works
(see~\cite{Feldman:2009wv, Holmes:2009uu, Feldman:2011me, Feldman:2010uv, Kane:2011tv, Roszkowski:2014wqa, Chakraborti:2015mra}). In the models considered here this connection is even stronger because the requirement that dark matter satisfy the WMAP and Planck relic density constraints forces the stop and neutralino masses to lie within $\sim30\GeV$ of each other for the lightest stops, and restricts the allowed parameter space of the model in the $m_0-m_{1/2}$ plane to a very narrow strip. Further, since $A_0/m_0$ is also determined to be in a very narrow range centered at $\sim-2.25$ in order to generate the right Higgs boson mass, the model is very predictive. It is of interest then to investigate the implications of these constraints for the direct detection of dark matter. For the parameter space of the model, the neutralino turns out to be almost exclusively a bino with  extremely small wino and Higgsino content. Therefore proton-neutralino scattering cross sections are expected to be small. The analysis of spin-independent and spin-dependent proton-neutralino cross sections is presented in table~\ref{tab:8}. The spin-independent cross sections lie in the range between $10^{-47}$ and $10^{-48}$ cm$^{-2}$. While cross sections of this size are decidedly small, they could still be visible in the next generation LUX-ZEPLIN (LZ) dark matter experiment, which is projected to reach a sensitivity of $\sim10^{-47}\text{cm}^{-2}$~\cite{Cushman:2013zza, Schumann:2015wfa}. Regarding the spin-dependent proton-neutralino cross section, the LUX-ZEPLIN will have a maximum sensitivity of $10^{-42}\text{cm}^{-2}$, which is still about three orders of magnitude smaller in sensitivity than what is needed to observe the spin-dependent proton-neutralino cross section exhibited in table~\ref{tab:8}. Thus the observation spin-dependent cross sections will be more difficult.

\section{Conclusion}\label{sec:conclusion}
In this work we have analyzed the lightest stop masses that can arise in high-scale models consistent with the Higgs boson mass constraint, relic density constraints from WMAP and Planck, and constraints on sparticle mass spectrum from RUN-I of the LHC. The specific focus of this work was mSUGRA, where it is found that stop masses as low as 400\GeV\, or lower can still exist consistent with these constraints. However, in the model space analyzed it is found that the stop must be the NLSP, with a mass $\sim1.1$ times the LSP mass. This restriction is needed in order to satisfy the relic density constraint via stop-neutralino coannihilation. 

{Requiring that the lightest neutralino supplies all of the thermal relic abundance then puts a lower bound on $\Delta m = m_{\tilde{t}_1} - m_{\tilde{\chi}_1^0}$ of approximately 30~GeV. Over the range of parameter space accessible in the lifetime of the LHC ({\em i.e.} within 3000 fb$^{-1}$ of integrated luminosity), this gap never exceeds 45\GeV. Thus neither the decay $\tilde{t}_1\to t\tilde{\chi}_1^0$ nor on-shell $W$-decays of the form $\tilde{t}_1 \to Wb\tilde{\chi}_1^0$ occur.} Rather the dominant decay of the stop in the allowed parameter space is the flavor-violating process $\tilde{t}_1\to c\tilde{\chi}^0_1$. 

{Production of superpartners is dominated by stop pair production, with all other processes reduced by approximately two orders of magnitude. The analysis of this work shows that the discovery of supersymetry for the class of models with light stops  discussed here can occur with an integrated luminosity as low as $\sim$ 60 fb$^{-1}$ at the LHC RUN-II, and will be dominated by searches involving low jet multiplicity, lepton vetoes, and large missing energy.} We have carried out a signature analysis for this class of models and find the dominant signature to be 2jl, where the characteristics of the 2jl SR are given in table~\ref{tab:3}. We have also carried out an analysis of the subdominant signature M1 {and an optimized signature 2jl-opt}, the characteristics of which are given in table~\ref{tab:3}. It is shown that the light stop can be discovered in 2jl with 88 fb$^{-1}$ {and in 2jl-opt with $\sim$ 60 fb $^{-1}$} of integrated luminosity at the LHC RUN-II, while a confirmation of the subdominant signal will require three times as much luminosity. {The minimum integrated luminosity can no doubt be reduced with further improvement in the efficiency to tag charm-jets~\cite{ATLAS:2014pla,Aad:2015gna,ATLAS2015}.}

We also discussed the dark matter associated with this class of models. It is shown that the spin-independent proton-neutralino cross section could be within reach of the next generation LUX-ZEPLIN dark matter detector while the spin-dependent proton-neutralino cross section will be more difficult to observe. {Any observed signal will be comparable to, but slightly larger than, coherent scattering by atmospheric neutrinos~\cite{Strigari:2009bq}.}

\textbf{Acknowledgments: }
We thank Darien Wood for a discussion.
This research  was  supported in part by the NSF grant PHY-1314774.

\end{document}